# Unravelling the mechanisms underlying crack initiation in additively manufactured steel

KenHee Ryou[a†], Yaozhong Zhang[a†], James A. D. Ball[b], Daniel Rubio-Ejchel[a], Dillon K. Jobes[a], Buhari Ibrahim[a], Allen Hunter[c], Charles Romain[d], Minh-Son Pham[e], Henry Proudhon[e], and Jerard V. Gordon[a*]

[a] Department of Mechanical Engineering, University of Michigan, Ann Arbor, MI 48109, USA

[b] ESRF, The European Synchrotron, 71 Avenue des Martyrs, CS40220, 38043 Grenoble Cedex 9, France

[c] Michigan Center for Materials Characterization, University of Michigan, Ann Arbor, MI 48109, USA

[d] ENSCM, Hétérochimie Moléculaire & Macromoléculaire 8, rue de l Ecole Normale, Montpellier, France

[e] Department of Materials, Imperial College London, Exhibition Road, South Kensington, SW7 2AZ, UK

[f] Mines Paris, PSL University, MAT – Centre des matériaux, CNRS UMR 7633, BP 87, 91003, Evry, France

* Corresponding author: Dr. Jerard V. Gordon (jerardvg@umich.edu) ; † Equal Contribution

## Abstract

Metal additive manufacturing (AM) is increasingly adopted for safety-critical applications across the biomedical, aerospace, and energy sectors. However, many AM alloys exhibit substantially lower fracture toughness and shorter fatigue lives than their wrought counterparts, limiting their structural reliability. The microscale mechanisms governing this deficit remain obscured because the evolution of crack-tip deformation cannot be directly resolved using conventional characterization techniques. Here, we combine in situ multimodal synchrotron X-ray diffraction with phase-contrast tomography to directly observe the three-dimensional evolution of crack-tip plasticity during loading. We find that wrought steel develops a localized crack-tip process zone characterized by extensive geometrically necessary dislocation (GND) accumulation, effective stress relaxation, and crack-tip blunting. This plastic zone provides shielding of the crack tip by redistributing deformation and reducing the local driving force for crack initiation. In contrast, the AM alloy exhibits suppressed GND evolution, limited crack-tip blunting, and persistent elevated stresses over an extended region ahead of the crack tip, indicating ineffective stress relaxation and premature crack initiation. These findings demonstrate that fracture resistance is governed by the spatial evolution of crack-tip plasticity, providing a mechanistic framework for improving the damage tolerance of AM structural alloys.

## Introduction

Laser powder bed fusion (PBF-LB) is the most widely adopted metal additive manufacturing (AM) technology, enabling the production of complex, high-performance components for aerospace, energy, biomedical, and nuclear applications. However, despite their high strength and the ability to produce near fully dense components, many PBF-LB alloys consistently exhibit lower fracture toughness[1–3] and shortened cyclic fatigue lives[1,2,4,5] compared to wrought counterparts, limiting their use in damage-tolerant structural applications. Although this behavior has traditionally been attributed to porosity and reduced ductility, many PBF-LB alloys continue to exhibit poor fracture resistance even in the absence of significant porosity and with excellent tensile ductility[3]. These observations suggest that the hierarchical deformation microstructures [1,2,6] produced by rapid solidification and repeated thermal cycling play a governing role, yet the underlying crack-tip mechanisms remain unresolved.

The origin of this reduction remains unresolved because multiple microstructural features evolve simultaneously during processing and deformation[7]. Porosity, reduced ductility, residual stresses, cellular microstructures, and oxide inclusions have all been proposed as contributors to reduced fracture resistance[2]. However, even in dense PBF-LB materials with relative densities exceeding 99.9%, significant reductions in fracture resistance are still observed[3,4], indicating that defect populations alone cannot explain crack initiation under these conditions. Similarly, PBF-LB alloys with excellent ductility such as 316L[8,9] still suffer the toughness reduction, suggesting that AM-characteristic microstructure and residual stresses are the main factors responsible for the toughness reduction. Fracture toughness reflects both the resistances to initiation of a crack and to the advanced of existing crack. Therefore, to understand the loss in toughness of AM alloys, it is important to study the effect of initial microstructure and stress and how they evolve and interact with other features (including pores).

One of the main reasons why it is challenging to unravel the origins of fracture resistance in AM alloys is that it requires direct measurement of how the crack-tip plasticity field evolves in relation to the surrounding microstructure inside solids. However, experimentally capturing crack initiation both spatially and temporarily remains challenging. Conventional fractography and *ex situ* characterization reveal final fracture morphology but cannot resolve the coupled evolution of crack geometry, stress redistribution, and microstructural deformation preceding crack initiation. Similarly, electron microscopy-based *in-situ* experiments (including electron backscattered diffraction) only offer surface observations. X-ray tomography provides three-dimensional crack morphology[10–14], while diffraction-based techniques quantify crystallographic deformation and internal stresses [15–18]. These approaches have largely been applied independently, leaving the coupled evolution of crack morphology, stress redistribution, and dislocation structures during crack initiation unresolved in PBF-LB alloys.

Here we present a detailed, comparative *in situ* study of PBF-LB and wrought 316L stainless steel specimens with nearly identical grain length scales and statistically equivalent low-porosity distributions. By designing samples with nearly identical grain and pore-size distributions, we isolate the role of the deformation microstructure in governing crack-tip plasticity. We use PBF-LB and wrought 316L stainless steel as a model system because 316L is widely used in structural, energy, chemical, biomedical, and nuclear applications, where fracture resistance is critical to reliability. In addition, PBF-LB processing of 316L produces high strength while retaining appreciable ductility (Supplementary S1), yet consistently reduced fracture resistance compared to wrought material, making it an ideal system for studying crack initiation under controlled microstructural conditions [2,3,19].

Using a multimodal *in-situ* synchrotron approach combining scanning three-dimensional X-ray diffraction (s3DXRD) and phase-contrast tomography (PCT), we simultaneously resolve crack morphology, residual stress fields, and microstructure (in particular, crystallographic orientation gradients, and geometrically necessary dislocation (GND) density at ~1 μm spatial resolution and track the evolution of these factors during loading[20]. We show that, despite nearly equivalent porosity and grain sizes, PBF-LB and wrought 316L develop fundamentally different crack-tip plasticity fields. The wrought alloy develops a localized crack-tip process zone characterized by steep GND and stress gradients that promote crack-tip blunting and shielding. In contrast, the PBF-LB alloy fails to develop this localized process zone, leaving elevated stresses distributed over a much larger region and promoting premature crack initiation. These results establish the spatial evolution of crack-tip plasticity as the governing microstructural mechanism controlling crack initiation in dense additively manufactured 316L stainless steel.

## Results

### *Initial microstructure and defect density*

Multimodal *in-situ* s3DXRD and PCT data acquired prior to loading reveal the initial microstructure and defect densities of both samples within the notch region (Methods). Initial s3DXRD datasets of multiple layers within the volume surrounding the notched region permit characterization of the grain structure prior to and after deformation. The PBF-LB specimen exhibits an irregular grain structure, showing a strong preferred crystallographic orientation in the <110> and <100> directions along the build axis (Fig. 1a and b). The dominance of these specific texture components is well known to occur in epitaxial solidification[21] during L-PBF processing of SS316L [22–25]. In contrast, the wrought specimen shows an equiaxed grain morphology and a more uniformly distributed crystallographic orientation as shown by the orientation spread in the inverse pole figure (IPF) map in Fig. 1c.

Statistical analysis of microstructural features in both samples is presented in Fig. 1d and e, showing the size distribution of grains and pores of both samples. The PBF-LB sample exhibits a similar grain size (median value of 15.1 μm grain diameter) compared to the wrought sample (median value of 17.5 μm). While the wrought sample contains grain boundaries consisting mainly of high-angle grain boundaries (44.9%) and twin boundaries (53.4%), with only 1.7% low-angle grain boundaries, the PBF-LB sample contains 52.7% low-angle grain boundaries (Fig. 1d). Crucially, both samples are highly dense with relative density of well above 99.98%, and porosity size distributions (including maximum pore sizes) being mostly identical statistically (Fig. 1e). By establishing that pore populations and grain length scales are statistically comparable between the two materials, we enable direct interrogation of how the initial deformation microstructure influences the evolution of crack-tip plasticity during loading.

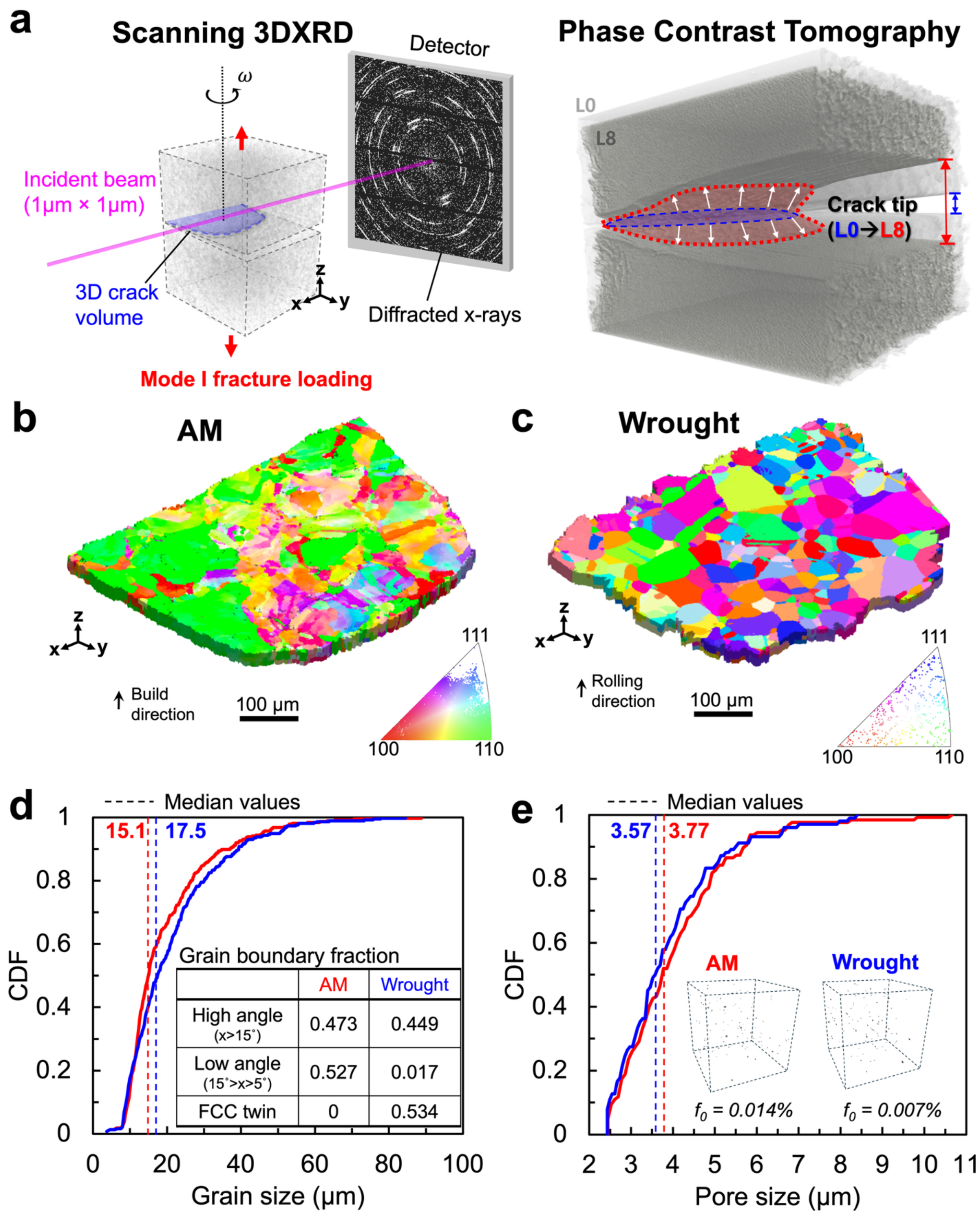

Grain boundary fraction

| | AM | Wrought |
|---|---|---|
| High angle (x>15°) | 0.473 | 0.449 |
| Low angle (15°>x>5°) | 0.527 | 0.017 |
| FCC twin | 0 | 0.534 |

**Fig. 1. Initial microstructure and defect density of PBF-LB and wrought SS316L**. (a) Schematic of the experimental approach combining synchrotron scanning 3D X-ray diffraction (s3DXRD) and phase-contrast tomography (PCT) to simultaneously measure crack morphology, intragranular stress fields, and crystallographic defect evolution during loading. Initial multilayer s3DXRD datasets with Inverse Pole Figure (IPF-Z) coloring for (b) PBF-LB and (c) Wrought

SS316L specimens. PBF-LB sample shows characteristic <110> texture along the build direction whereas wrought sample displaying an equiaxed grain morphology. (d) Cumulative distribution function (CDF) comparing grain size (area) and grain boundary character between the two material conditions. (e) Comparative pore size distributions indicating statistically similar defect populations in both PBF-LB and Wrought samples.

***Evolution of crack tip morphology and fracture resistance***

To quantify the evolution of the fracture driving force during loading toward crack initiation, crack tip opening displacement values (CTOD) were calculated from *in-situ* PCT data (Supplementary S2 and S3) to quantify local fracture resistance in PBF-LB and wrought at given load steps. CTOD measurements in Fig. 2a reveal that the wrought sample showed a stable increase in CTOD with crack extension, indicating a progressive increase in resistance to crack advance consistent with increasing crack-tip plastic accommodation[26]. This is evidenced by the significant increase from 3.6 μm at load step 3 (L3) to a maximum value of 59.2 μm at load step 8 (L8) over a relatively small change in crack length of 15 μm. In contrast, the PBF-LB sample only provides a modest increase of CTOD, from a value of 2.6 μm to a maximum value of 13.6 μm at L6 (Fig. 2a) which is 58.3% lower than the wrought samples at similar loads. These results echo recent review articles by Paul et. al [3] and Zhu and Chen[4] which demonstrate that fracture toughness of as-manufactured PBF-LB Fe, Ti, and Al alloys can drop by as much as 70% compared to their wrought counterparts [27]. Given the nearly identical porosity condition between the PBF-LB and wrought, this data suggests that a lower resistance to crack advance of the PBF-LB material is not mainly due to the presence of porosity.

To further interrogate the origins of the reduced crack resistance in the PBF-LB microstructure, we employed *in situ* 3D PCT data to directly track the development of crack, including initiation and interactions between the growing crack and pre-existing processing defects[12,28–30] (Fig. 2c and 2d). Both samples show disparate development of the crack-tip morphology. The wrought specimen develops a progressively blunted crack tip throughout loading, indicating sustained plastic accommodation prior to unstable crack extension (Fig. 2c). Even when the crack encounters pre-existing pore (diameter of 6μm), this crack tip blunting sustained; as the crack front reaches the pore, it integrates the defect without forming a sharp stress concentrator, preserving a rounded tip profile (Fig. 2c). Furthermore, 3D observations reveal that the pore enlarges during deformation and becomes noticeably elongated along the loading axis (Fig. 2c).

In contrast, the PBF-LB microstructure demonstrates significant spatial irregularities in the crack front retreat. Instead of blunting, the crack tip underwent a localized transition to a sharp fracture mode (Fig. 2d). When interacting with an internal process-induced pore (likely an entrapped gas void[7] of 5 μm in diameter, i.e. almost the same pore size in the wrought case), a sharp crack tip developed and rapidly linked with the neighboring pore, bypassing the stabilizing effect of plastic blunting. Notably, the 3D reconstruction confirms that while the pore enlarged during deformation, it did not elongate along the loading axis in the PBF-LB sample. The absence of pore elongation indicates that the surrounding matrix cannot develop the large crack-tip plastic deformation required for stable void growth[31,32]. These observations demonstrate that, when defect densities remain low, crack-defect interactions are governed primarily by the propensity of crack tip blunting thanks to local plasticity rather than the defects themselves[33].

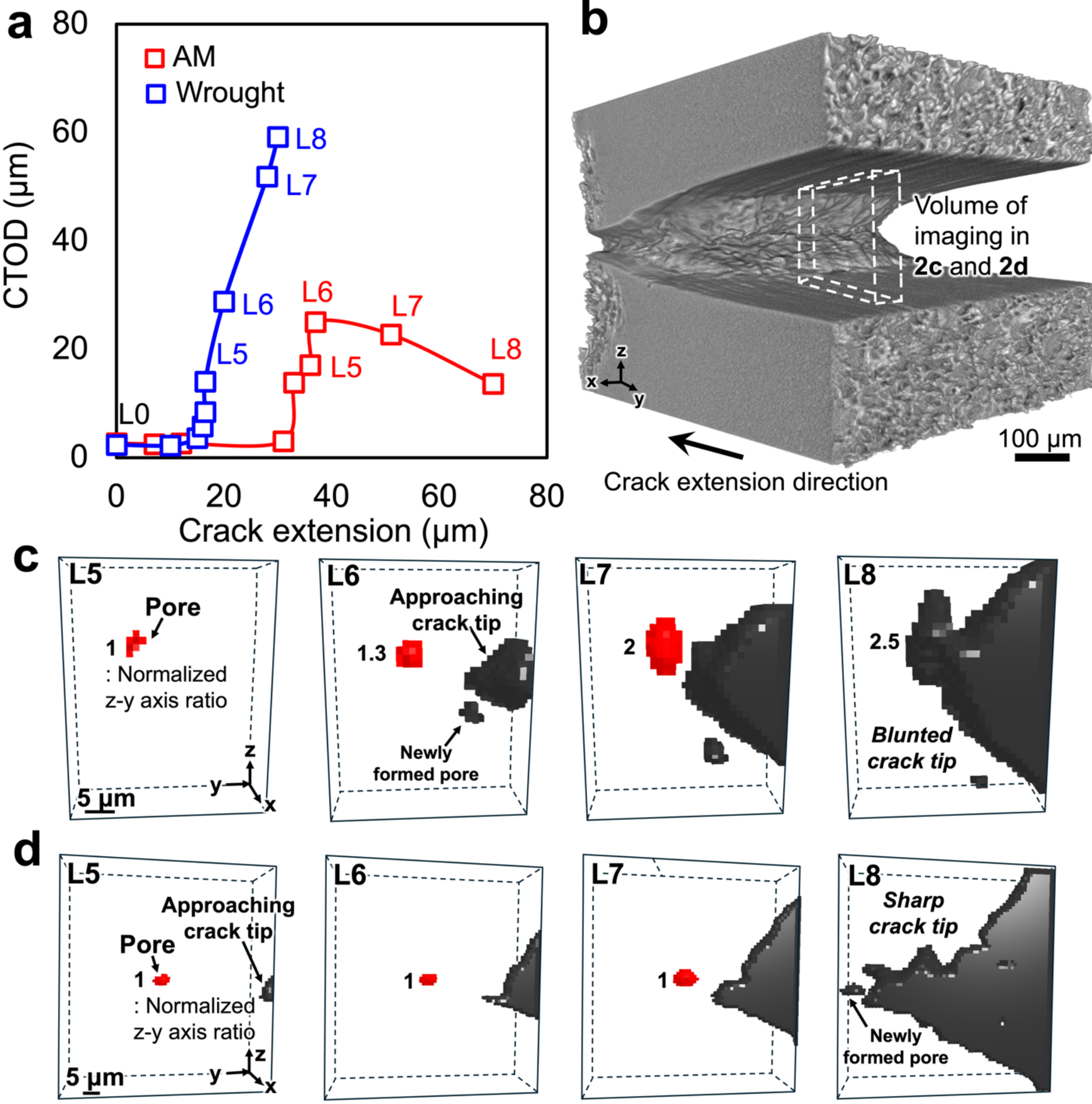

**Fig. 2. Crack initiation and mesoscale fracture response**. (a) 3D PCT data was used to measure crack tip opening displacement (CTOD) versus against crack extension. (b) Schematic of bulk 3D volume used for imaging of crack tip morphology in 2c and 2d. (c) The wrought alloy shows pronounced crack-tip blunting and more distributed plastic deformation ahead of the crack tip during steady state crack advance and interaction with a pore defect. (d) The PBF-LB alloy exhibits reduced crack-tip blunting after interactions with processing pores.

### *GND evolution during crack opening*

To understand the microstructural origins driving the differences in fracture resistance, geometrically necessary dislocation (GND) density ($\rho_{GND}$) was quantified for both samples throughout loading via *in-situ* s3DXRD data (Fig. 3). This analysis provided a direct measure of the local lattice curvature that is due to the residual stress after fabrication (i.e. initial state) and later required to maintain plastic compatibility and provide direct visualization of plastic strain gradients at the crack tip during deformation.

A striking disparity is immediately evident in the initial (L0) state. The PBF-LB material exhibits a high baseline dislocation density (~$10^{15}$ $m^{-2}$) in region ahead of crack tip (Fig. 3a), consistent with a high residual stress that is expected for a PBF-LB condition and results from the strong thermal gradient and severe thermal cycles inherent to the printing process[7]. In contrast, the wrought baseline shows a negligible initial GND population (Fig. 3b), characteristic of almost residual stress-free condition, typical of a fully annealed structure.

Under loading, the wrought alloy develops a progressively expanding GND field extending ahead of the crack tip (Fig. 3b and c). The increasing spatial gradient indicates continuous plastic accommodation throughout the process zone rather than localized deformation immediately at the crack tip, effectively blunting the crack tip and strengthening the material to prevent a plastic localization ahead of the crack tip [15,34–36]. In addition, GND density considerably dropped off further way from the crack tip in the wrought, while it remained almost unchanged in the AM condition, Fig. 3d. The steep GND gradient ahead of the crack tip in the wrought provides soft process zone that lowers driving force for the crack advance, i.e. the crack tip shielding. Conversely, the PBF-LB sample already exhibits an initially high accompanied by a very limited increase of $\rho_{GND}$ (nearly uniform magnitude), indicating that limited plasticity developed in front of the crack tip (i.e. much less crack tip blunting). In addition, a relatively unchanged profile of GNDs further away from the crack tip indicates that the high driving force for crack advance remained and extended in region ahead of crack tip (Fig. 3d).

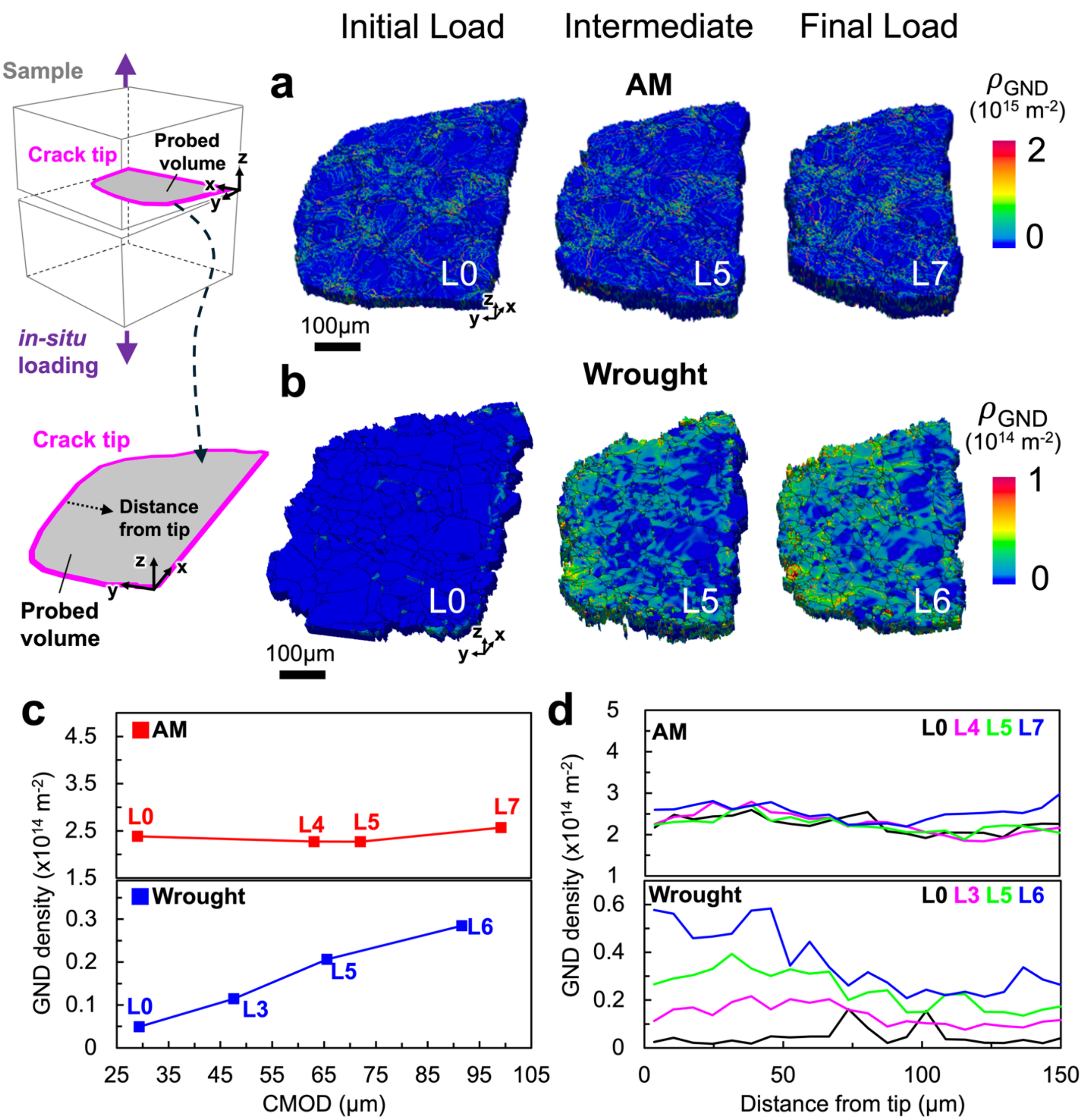

**Fig. 3. Spatial distributions of GND density during loading.** Spatial distributions of GND density reconstructed from voxelized s3DXRD data for PBF-LB and wrought specimens at equivalent load steps ("L#"). (a) The PBF-LB microstructure shows GND stagnation whereas (b) average wrought values show steady increase of GND with crack mouth opening displacement (CMOD). Spatial profiles of GND from the crack tip show (c) a uniform plateau in PBF-LB sample indicating limited plastic accommodation versus (d) a steep decay from the notch tip in the wrought microstructures during loading.

***Evolution of local intragranular stress fields***

To quantify the local residual stress evolution and understand the mechanical consequences of limited GND accumulation in the PBF-LB material, intragranular von Mises stress ($\sigma_{VM}$) fields were experimentally quantified during crack opening using voxelized s3DXRD measurements[37] (Supplementary S4). Fig. 4a shows the evolution of volume-averaged intragranular von Mises stress as a function of crack mouth opening displacement (CMOD; Supplementary S2) for both PBF-LB and wrought conditions. Both specimens exhibited an initial average stress state of approximately 200 MPa at L0 due to the applied seating preload required for stable specimen alignment within the loading device (Methods). The magnitude and spatial heterogeneity of the initial intragranular stresses, particularly in the PBF-LB material, are consistent with our previous s3DXRD measurements of independent as-built PBF-LB 316L specimens[20], which demonstrated persistent grain-scale residual stresses generated by solidification-induced microstructural heterogeneity. During subsequent loading, the stress evolution diverges markedly between the two conditions. The wrought sample exhibits only a modest increase in average stress, reaching 355 MPa at L6, whereas the PBF-LB sample increases to approximately 800 MPa at L7, indicating substantial internal stress accumulation and limited stress relaxation (Fig. 4b,c).

One-dimensional stress profiles also reveal fundamentally different stress fields surrounding the crack tip. In the wrought alloy, von Mises stress decreases rapidly (by ~100 MPa) over a distance of 75 μm from the crack tip (Fig. 4c and i), consistent with a drop off in GNDs shown in Fig. 3b and d, resulting in efficient stress relaxation within a crack-tip plastic zone[38], i.e. shielding effect. In contrast, elevated stresses remain high and extend across nearly the entire 125 μm region surrounding the crack tip in the PBF-LB alloy (Fig. 4b), indicating inefficient stress relaxation and suggesting that the surrounding material does not participate in plastic accommodation to locally relieve elevated stress, i.e. the crack tip shielding is negligible[39].

To probe the effect of local microstructure on internal stress field evolution and plastic accommodation, representative grains were selected from high-stress (>600 MPa) and low-stress (<400 MPa) regions of the crack-tip process zone (Fig. 4e and h). In the PBF-LB material, grains carrying the highest retained stresses generally correspond to hard orientations that are less favorable for slip activation, whereas lower-stress grains correspond to orientations more conducive to plastic deformation[40]. For the two conditions, hard grains were found to have their [113] or [103] close to the loading direction (LD), while those with [101] close to LD were soft (Fig. 4e and f). The stress contrast between these populations is significantly larger in PBF-LB than in the wrought alloy, indicating stronger plastic inhomogeneity in the PBF-LB microstructure[41].

Together, the GND and stress fields reveal that the wrought alloy develops a relatively homogeneous plastic zone over which the material hardens isotopically to prevent plastic localization. This homogeneous plasticity promotes crack-tip blunting. In addition, the plastic zone is surrounded by much softer regions of much lower GNDs and stresses (Fig. 3d and Fig. 4c), resulting in considerable shielding. In contrast, the PBF-LB alloy fails to develop this process zone. Instead, elevated stresses persist over a much broader region ahead of the crack tip, i.e. extending high driving forces for crack advance that results in negligible crack-tip shielding. In addition, it promotes stress partitioning in preferentially oriented grains, leading to more plastic localization that prevents the crack tip blunting.

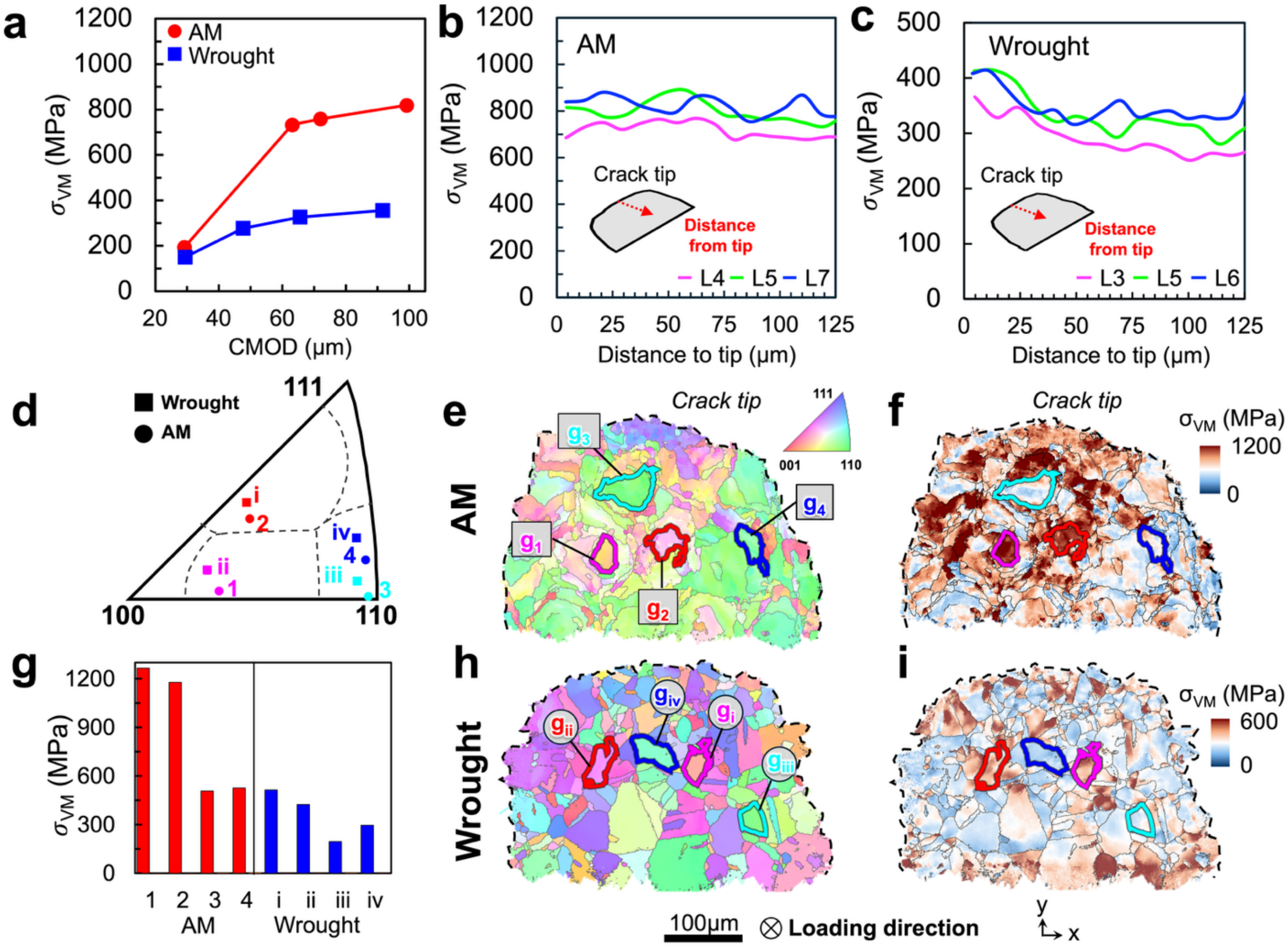

**Fig. 4. Evolution of intragranular stress fields and grain-scale partitioning**. (a) Volume-averaged intragranular von Mises stress as a function of crack opening displacement for PBF-LB and wrought alloys. 1D line profiles demonstrate rapid stress decay away from the crack tip in the wrought alloy, whereas elevated stresses persist over substantially larger distances in the PBF-LB alloy (b-c). Intragranular stress partitioning across selected grains (d) with different crystallographic orientations for (e) PBF-LB and (h) wrought samples (Note: (e) and (h) are inverse pole figures along the loading direction). These grains were in high- and low-stress regions of the crack-tip process zone with (f) PBF-LB specimen showing high stress retention (>1200 MPa) and localized hotspots whereas (i) wrought specimen showing effective stress relaxation and homogeneous stress distribution.

## Discussion

The multimodal *in situ* measurements provide a direct three-dimensional view of crack-tip deformation, revealing that the spatial evolution of the crack-tip plasticity field governs the blunting and shielding of crack tip that, in turn, affect crack initiation in PBF-LB 316L. The key distinction between the two material conditions is not simply their initial states of dislocation density, pore volume fractions, or residual stress distributions, but the evolution of microstructure both temporarily and spatially influence in the development of plastic deformation ahead of the crack tip during loading. The wrought alloy develops a relatively homogeneous crack-tip plastic zone thanks to considerable generation of dislocations (e.g. GNDs, Fig. 3), that result in substantial crack tip blunting even when it interacts with an existing pore (Fig. 2a and c and Fig. S2). Such a blunting effect makes the influence of microstructure much stronger than that of process defects such as pores. The increase in density and homogeneous distribution of dislocations lead to high hardening rate locally in front of crack tip, preventing localization upon further loading. In addition, beyond the plastic zone, the material is relatively soft reflected by a quick drop in GNDs (Fig. 3d) and stress (Fig. 4c). This soft buffering zone efficiently shields the crack tip. Both the blunting and shielding help to considerably delay crack initiation and slow down the propagation of micro-short cracks.

In contrast, the PBF-LB alloy fails to develop this homogeneous process zone due to exhaustion in dislocation generation, leading to insignificant crack tip blunting (Fig. 2a and d, and Fig. S3). In addition, elevated internal stresses extended over a much larger region ahead of the crack tip. Negligible reduction of dislocations and stress ahead of crack tip, resulting in no shielding effect. Moreover, plasticity, if developed in the crack tip region, is only localized in a group of "soft" grains, causing deformation highly heterogeneous (Fig. 4d, e and f). Even in these soft grains, dislocation increase is still marginal, leading to insignificant hardening in these soft grains. In other words, the PBF-LB alloy exhibits a markedly reduced capacity to develop an effective crack-tip process zone, promoting plastic localization, earlier crack initiation, and limited crack-tip blunting compared with the wrought condition.[42]

More broadly, our findings not only provide more accurate interpretation of fracture behavior but also offer a physical basis in designing alloys and process to enhance the damage tolerance for PBF-LB metallic components. While the present study employs monotonic loading, the observed inability to develop a localized crack-tip plasticity field has direct implications for high-cycle fatigue (HCF), where lifetime is dominated by crack initiation and early-stage growth [2,5,43,44]. In such regimes, microstructure is highly influential[6] and the ability of microstructure to blunt and shield the crack tip through dislocation-induced hardening and soft buffering zone is essential for delaying crack nucleation and propagation. The present results suggest that when dislocation generation is marginal, crack tip plasticity remains limited and only localized in separated grains of soft orientations, plastic localization develops more rapidly, promoting earlier crack initiation and faster advance under cyclic loading. Direct extrapolation to cyclic deformation requires caution because reversible slip and evolving dislocation structures will modify crack-tip plasticity. Future *in situ* studies on both AM *in situ* and *ex situ* thermally annealed or otherwise microstructurally modified PBF-LB alloys will be important for isolating the role of the cellular dislocation architecture in governing crack initiation resistance under both monotonic and cyclic loading.

These findings suggest that improving fracture resistance in AM alloys will require controlling not only defect populations (including porosity and initial dislocation density) in fabrication but also the spatial evolution of microstructure and how it affects the development of crack-tip plasticity. Processing routes that promote the formation of a well-defined and homogeneous crack-tip plastic zone through engineering the AM-characteristic cellular dislocation structures may therefore provide a pathway for simultaneously improving fracture and fatigue resistance in AM [45–47]. More broadly, these results identify the spatial evolution of dislocations and associated crack-tip plasticity field as a fundamental microstructural descriptor governing crack initiation in AM, providing a mechanistic framework for designing alloys and optimizing the AM processes to achieve metallic components with improved damage tolerance.

## Methods

### *Sample fabrication*

Two samples were investigated: (i) as-built PBF-LB SS316L and (ii) conventionally wrought and heat-treated SS316L. PBF-LB SS316L specimens were fabricated using pre-alloyed gas atomized powder on an EOS M290 system using manufacturer-recommended parameters (195 W laser power, 1083 mm/s scanning velocity, 90 μm hatch spacing, 20 μm layer thickness, and 67° interlayer rotation). Wrought SS316L specimens were machined from hot-rolled plate and solution annealed at 1050°C for 30 min followed by rapid quenching to obtain a fully austenitic equiaxed microstructure.

Sub-sized Nanox tensile specimens were extracted from both materials using wire EDM, with the loading direction aligned parallel to the build direction for PBF-LB and the rolling direction for wrought SS316L. This. specimen geometry was selected to enable high-energy X-ray transmission and full-field crack-tip characterization. A sharp mode-I notch with depth of ~200μm was introduced using plasma FIB milling to provide a well-defined crack initiation site.

*In-situ synchrotron fracture experiments*

*In situ* fracture experiments were performed at ESRF beamline ID11 using a Nanox miniature loading device [48]. A 43.57 keV monochromatic X-ray beam was used for simultaneous phase-contrast tomography (PCT) and scanning 3D X-ray diffraction (s3DXRD) measurements during incremental loading. PCT was acquired at 0.93 μm voxel resolution to capture crack morphology, crack-tip blunting, and internal defects. s3DXRD measurements were performed within a 500 × 350 × 50 μm³ volume centered at the crack tip using a 1.02 × 1.3 μm beam, enabling grain-resolved orientation and elastic strain measurements during fracture evolution.

A small preload of 3–7 N was applied prior to the first measurement at load step zero (L0) to ensure proper seating within the loading grips while maintaining an elastic baseline condition. This preload corresponds to nominal stresses below 20% of the respective yield strengths and does not introduce measurable plastic deformation prior to crack-tip measurements (Fig. S4a,b). The resulting L0 stress state therefore represents the combined contribution of the applied elastic preload and pre-existing grain-scale residual stresses. To enable direct comparison between materials with different yield strengths, subsequent measurements were synchronized using crack mouth opening displacement (CMOD) values extracted from PCT reconstructions (Supplementary Table 1). Measurements were acquired following partial unloading to minimize stress relaxation effects.

*PCT and s3DXRD reconstruction*

PCT data were reconstructed using the Nabu software package[49]. s3DXRD data were reconstructed using ImageD11[50], following established procedures [20] (Supplementary). Grain orientations and elastic strain tensors were obtained through voxel-resolved diffraction peak indexing and refinement. These measurements enabled calculation of intragranular stresses and geometrically necessary dislocation (GND) densities near the crack tip. The PCT and s3DXRD reconstructed volumes were registered using common surface features and notch geometry to correlate crack morphology with grain-scale deformation fields.

*Crack tip displacement, stress and, GND calculations*

Crack-tip opening displacement (CTOD) was quantified from reconstructed PCT volumes following ASTM E1820 definitions. CTOD values were used as a comparative measure of crack resistance between microstructures rather than absolute plane-strain fracture toughness values because of the sub-sized specimen geometry. GND density was quantified using the Nye tensor method [51] from voxelized orientation fields processed using MTEX[51]. Intragranular stresses were calculated from measured elastic strain tensors using the anisotropic elastic stiffness tensor of SS316L[37] (Supplementary S4).

## Data availability

S3DXRD and PCT datasets supporting this study are deposited at the ESRF data repository and are accessible at https://doi.org/10.15151/ESRF-ES-2169731953. Source data for all figures are available from the corresponding author upon reasonable request.

## Code availability

S3DXRD reconstruction was performed using ImageD11 (https://github.com/FABLE-3DXRD/ImageD11). PCT reconstruction used the Nabu package (ESSRF). Custom analysis scripts are available from the corresponding author upon reasonable request.

## Competing interests

The authors declare no competing interests.

## Author Contributions

K. R.: Formal analysis, Writing – original draft, Writing – review & editing. Y. Z.: Formal analysis, Software, Writing – original draft. J.A. D. B.: Investigation, Formal analysis, Software. D. R.: Investigation. D. J.: Investigation. B. I.: Investigation. A. H.: Investigation. C. R.: Investigation. H. P.: Supervision, Resources, Investigation, Funding acquisition. J.V.G: Conceptualization, Investigation, Writing – original draft, Funding acquisition, Writing – review & editing, Project administration.

## Acknowledgements

We are grateful to Dr. Philip J. Noell (Sandia National Laboratories) for insightful comments and feedback on an early draft of this manuscript. We are grateful to the ESRF for the provision of beamtime on ID11 under proposal numbers MA-6524. This work was partially supported by NSF DMR- 2237433 and the DOE Office of Nuclear Energy's Nuclear Energy University Programs (NEUP), under Contract DE-NE0009413. The authors acknowledge the financial support of the University of Michigan College of Engineering and NSF grant #DMR-0320740, and technical support from the Michigan Center for Materials Characterization.